\documentclass[epj]{webofc}
\usepackage[varg]{txfonts}   
\woctitle{Connecting The Dots 2016}
\newcommand{\Kalman}{K\'alm\'an\xspace}
\begin{document}
\title{A New Track Reconstruction Algorithm suitable for Parallel Processing
based on Hit Triplets and Broken Lines}
%
%

\author{\firstname{Andr\'e}
  \lastname{Sch\"oning}\inst{1}\fnsep\thanks{\email{schoning@physi.uni-heidelberg.de}}
}

\institute{Physikalisches Institut, Ruprecht-Karls Univerisit\"at Heidelberg,
  Heidelberg, Germany 
          }

\abstract{%
Track reconstruction in high track multiplicity environments at current and
future high rate particle physics experiments is a big challenge and very time
consuming. 
The search for track seeds and the fitting of track
candidates are usually the most time consuming steps in the track reconstruction.
Here, a new and fast track reconstruction method based on hit triplets is proposed which
exploits a three-dimensional fit model including multiple scattering and 
hit uncertainties from the very start, including the search for track seeds. 
The hit triplet based reconstruction method assumes a 
homogeneous magnetic field which allows to give an analytical solutions for
the triplet fit result. 
This method is highly parallelizable, needs fewer operations 
than other standard track reconstruction methods 
and is therefore ideal for the implementation on parallel computing
architectures.
The proposed track reconstruction algorithm has been studied in the context of
the Mu3e-experiment and a typical LHC experiment.
}
\maketitle
\section{Introduction}
\label{intro}
Track reconstruction in high energy experiments at high luminosity colliders 
is a big challenge.
The large particle rates together with the track
extrapolation uncertainties, generated by multiple Coulomb scattering in the
silicon tracking detector material, lead to hit confusion and define
the hit combinatorial problem.
High track multiplicities not only increase considerably the computational
effort for solving this combinatorial problem, they also reduce the track finding
efficiency and purity.

Standard track reconstruction algorithms employed in high track multiplicity environments 
consist typically of the following  steps:
\begin{itemize} 
\item[1] {\bf Seed} search: A minimum number of hits in several tracking layers
  is required to form a track seed. 
  For the reconstruction of initial track parameters at least three
  points are required. 
  This could be hits in three detector layers or in only two detector layers 
  if additional constraints on the origin of the track, e.g. collision
  point, are employed.
\item[2] {\bf Fitting} of track prototypes and seeds: 
  By using a tracking model, which ideally should include spatial hit uncertainties as well as
  multiple scattering uncertainties, 
  track parameters of seeds and prototypes\footnote{
  A track prototype is defined to be an incomplete track candidate which has
  usually more than three hits.} are determined. 
\item[3] {\bf Hit linking}: Using previously fitted track parameters, 
  the particle trajectory of the track prototype is extrapolated to link hits
  from further detector layers. 
  Track extrapolation uncertainties are usually included by defining a search window.
  Hit linking is terminated if hits  in all tracking layers are found or no
  further hits can be found because of missing hits.
\item[4] {\bf Fit quality:} Track candidates have to fulfill some quality
  criteria, that is a minimum number of linked hits (layers), minimum track
  length, minimum track fit  quality, etc.  
\item[5] {\bf Arbitration} between track candidates: Different track candidates can
  share one or several hits. An arbitration algorithm can be used to find
  the ``best'' track candidate. Other track candidates sharing the same hits are discarded.
\end{itemize} 
The steps 2 and 3 are iterated until all hits are linked or no more track
candidates can be found. 
The \Kalman filter~\cite{Kalman:1960:nal, Fruhwirth:1987:akf, BilloirEtAl:1990:spr} 
 is an example for such an iterative algorithm which is 
used as standard tool for track reconstruction in many particle physics
experiments.

Besides track fitting the most time consuming step is usually the 
track seed finding due to the hit combinatorics.
For a track seed search in $n_{layer}$ layers the processing time
scales with $(N_{\rm hit})^{n_{layer}}$ where $N_{\rm hit}$ is the number of
hits. 
For this reason track seed searches are often performed in regions
of low occupancies (i.e. outer tracking layers in case of collider experiments).
Different methods can 
be used to further reduce the hit combinatorial problem in the
track seed search in the successive hit linking step: 
\begin{itemize}
\item application of simple geometrical preselection cuts,
\item application of more advanced algorithms like cellular automata~\cite{Abt:2002he},
\item Hough transforms~\cite{ref:Hough},
\item usage of lookup tables or associative memories.
\end{itemize}

Here, a new method is proposed for solving the hit combinatorial problem for
seed finding and linking of hits to tracks. 
It can be used to accelerate the steps 1 -- 4 in comparison with other standard
reconstruction algorithms.\footnote{The arbitration step is not addressed further here.}
The proposed
method is based on hit triplets for which the track parameters can be instantly
calculated by exploiting a three-dimensional fit model including multiple scattering and 
hit uncertainties. 
This fit requires the three-dimensional hit coordinates including their uncertainties
and is only applicable in homogeneous magnetic fields.
Both requirements are usually fulfilled in modern pixel tracking devices.

\section{Method}
\label{sec-1}

\begin{figure}[t]
\centering
\includegraphics[width=6cm,clip]{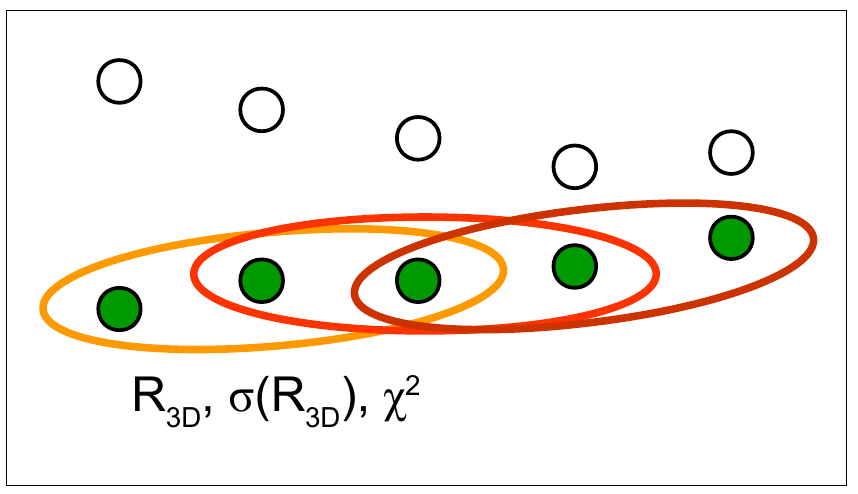}
\caption{Sketch of merging hit triplets to full tracks. The track fit result
  of each triplet is given by the three-dimensional radius $R_{3D}$, its
  uncertainty $\sigma(R_{3D})$ and the fit quality $\chi^2$.}
\label{fig:triplet_merging}       
\end{figure}

The hit triplet based track reconstruction method works as follows.
In a first step hit triplets are found by fitting three-hit combinations and
by applying a $\chi^2$-cut on the triplet fit quality. 
What is new compared with other similar methods is here the full exploitation of the tracking
model at this very first track reconstruction step, thus reducing the number
of wrong seeds and mitigating the combinatorial problem.
As this method is rigorous it also works for ``large triplets'' with far distant hits and large
track curvatures.
After having found valid hit triplets, conjoined triplets
(triplets which share two hits) are merged successively to full tracks, 
see~Figure~\ref{fig:triplet_merging}. 
The resulting trajectory is finally represented by broken lines. 
The track parameters are combined using a simple and fast sum
which is explained in more detail in Section~\ref{sec:merging}.
This fast summing technique is much faster than other reconstruction methods
which are based on iterations and rely on matrix inversions. 
Furthermore, the fast summing technique is exact if multiple scattering uncertainties
dominate, being the case for the majority of particles at hadron colliders if
silicon tracking detectors are used.

Above described merging technique only includes local but no long-range
hit correlations which are important for the reconstruction of high momentum tracks.
For high momentum tracks spatial hit uncertainties dominate over multiple
scattering uncertainties and a global fit is required.
This can be realized by applying a General Broken Line Fit 
(GBL)~\cite{Blobel:2006:nft, BlobelEtAl:2011:fac, Kleinwort:2012:gbl}
after triplet merging.
The GBL fit is an extended track fit that takes into account both,
scattering effects and spatial uncertainties, and has been shown
\cite{Kleinwort:2012:gbl} to be equivalent to the \Kalman
filter~\cite{Fruhwirth:1987:akf}.
The GBL algorithm performs a linearized fit by varying positions and kink angles 
at selected points around a reference trajectory, which is here chosen to be
the merged triplet trajectory.

\subsection{Fitting of Hit Triplets}

\begin{figure}[t!]
\centering 
\includegraphics[width=0.75\textwidth]{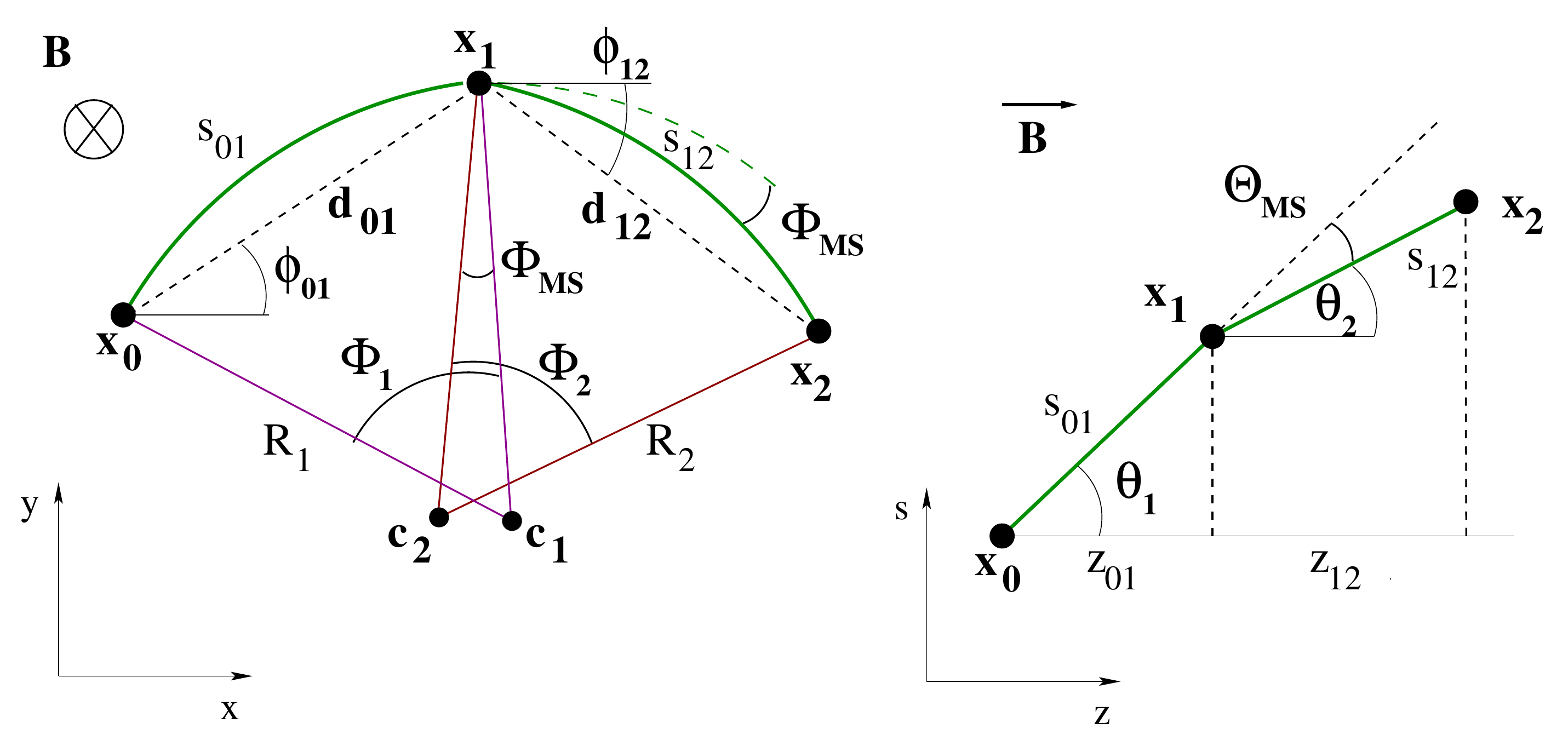}
\caption{Particle trajectory in a homogeneous magnetic field defined
  by a triplet of hits $\mathbf{x_0}$, $\mathbf{x_1}$ and
  $\mathbf{x_2}$.  The left view shows
  the projection to the plane transverse to the magnetic field,
  whereas the right view is a projection to the field axis-arclength
  ($s$) plane.  
  $\Phi_{MS}$ and $\Theta_{MS}$ denote the multiple scattering angles
  in the transverse and longitudinal plane. 
}
\label{fig:triplet}
\end{figure}

The most general objective function for track fitting includes multiple
scattering and spatial hit uncertainties
\begin{equation}
\chi^2 \ = \ \frac{\Theta_{MS}^2}{\sigma_\theta^2} + \frac{\Phi_{MS}^2}{\sigma_\phi^2}
+ \sum_{jk} ({\bf x_j}-{\bf \xi_j}) {\bf V^{-1}_{jk}} ({\bf x_k}-{\bf \xi_k})
\nonumber
\ .
\label{eq:definition_chi2}
\end{equation}
Here, $\Phi_{MS}$ and $\Theta_{MS}$ are the multiple scattering
uncertainties in the transverse and longitudinal plane, see Figure~\ref{fig:triplet}. 
${\bf x_k}$ and ${\bf \xi_k}$ are the fitted and measured hit coordinates and
${\bf V}$ is the covariance matrix.

For a particle moving in a homogeneous magnetic field an analytical solution
can be calculated for a trajectory measured at three points (hit triplet). 
The solution for the case of negligible spatial hit uncertainties is briefly
summarized in the next section (for a full description see~\cite{ref:triplet_paper}).
The solution for including spatial hit
uncertainties is discussed in the following Section~\ref{sec:triplet_general}.

\subsubsection{Triplet Fit with Multiple Scattering}

If spatial uncertainties are negligible, i.e. ${\bf x_k}={\bf \xi_k}$,
the right side of
Equation~\ref{eq:definition_chi2} can be neglected.
Assuming energy conservation the fit has one degree of
freedom.\footnote{Energy loss in silicon detectors is typically small. 
If required energy loss can also be
  corrected for in the triplet fit procedure.} 
The multiple scattering angles projected in the transverse and longitudinal
plane can then be written as a function of the
three-dimensional radius $\Phi_{MS}=\Phi_{MS}(R_{3D})$ and
$\Theta_{MS}=\Theta_{MS}(R_{3D})$ which in turn is related to the momentum 
$p$ of the particle via
$R_{3D}=p/(qB)$ with $q$ being the electric charge.  
$\Phi_{MS}(R_{3D})$ and $\Theta_{MS}(R_{3D})$ are
transcendent functions which can not be algebraically solved. 
However, the minimization condition
\label{sec:tripletms}
\begin{equation}
\frac{{d}}{{d} R_{{3D}}} \chi^2(R_{{3D}}) =0
\nonumber
\end{equation}
can be solved by linearization around the circle solution in the
transverse plane which is described in detail in~\cite{ref:triplet_paper}.  
The result\footnote{
In Reference~\cite{ref:triplet_paper} a second formula 
for $R_{{3D}}$ is given which includes a bias correction.
For sake of  simplicity we use here the result 
from Equation~23 in~\cite{ref:triplet_paper}.
}
 for the three-dimensional radius
\begin{equation}
R_{{3D}}  =  - \; \frac{\eta \; \tilde{\Phi}_{C} \; \sin^2{\vartheta_C}  + \beta
  \; \tilde{\Theta}_{C}  }{\eta^2 \; \sin^2{\vartheta_C}  + \beta^2}
 \quad 
\label{eq:r3d}
\end{equation}
depends only on parameters derived from the relative position of the hit
coordinates, $\eta, \tilde{\Phi}_{C}, \beta, \tilde{\Theta}_{C}=f({\bf x_k})$,
and the polar angle $\vartheta_C$ which can be approximated  from the circle 
solution. 
The exact definitions of these geometric parameters can be found 
in~\cite{ref:triplet_paper}.
The corresponding uncertainty is given by
\begin{equation}
\sigma({R_{3D}})  \ = \ \sigma_{MS} \; \sqrt{\frac{1}{\eta^2 \; \sin^2{\vartheta_C}+\beta^2}} 
\nonumber
\label{eq:sig_R}
\end{equation}
and contains the expected multiple scattering uncertainty, 
which can be calculated using the Highland approximation~\cite{Highland:1975:prm}
\begin{equation}
 \sigma_{MS} \approx \frac{13.6~\textrm{MeV/c}}{\beta p} \sqrt{X/X_0}
\label{eq:highland}
\end{equation}
after fitting, using the effective material thickness $X$ at the middle hit
position where the scattering takes place. 
Finally the fit quality is given by
\begin{equation}
\chi_{min}^2  =
\frac{1}{\sigma_{MS}^2} \ 
\frac{(\beta \; \tilde{\Phi}_{C}  - \eta \; \tilde{\Theta}_{C})^2}{
  \eta^2 + \beta^2  / \sin^2{\vartheta_C}} \ .
\nonumber
\label{eq:chi2}
\end{equation}

\subsubsection{Triplet Fit with Hit Uncertainties}
\label{sec:triplet_general}
In case of non-negligible spatial hit uncertainties the above description
yields too large $\chi^2$-values and will fail. 
For hit triplets this bias can be fixed elegantly 
 by absorbing the hit uncertainties in the scattering angles.
By introducing effective scattering angles
\begin{equation}
{\Delta}\Theta \ = \ \Theta_{\rm MS} + \Theta_{\rm hit} 
\quad \textrm{and} \quad 
{\Delta}\Phi \ = \ \Phi_{\rm MS} + \Phi_{\rm hit} \ ,
 \nonumber
\end{equation} 
a new objective function can be defined
\begin{equation}
\chi^2 \ = \ \frac{{\Delta}\Theta^2}{{\sigma_\theta}^2_\textrm{eff}} +
\frac{{\Delta} \Phi^2}{{\sigma_\phi^2}_\textrm{eff}} ,
 \nonumber
\end{equation}
where effective hit uncertainties have been introduced
\begin{equation}
{\sigma_{\theta}^2}_\textrm{eff}=\sigma_{\rm MS}^2+\sigma_{\theta_{hit}}^2
\quad \textrm{and} \quad 
{\sigma_{\Phi}^2}_\textrm{eff}=\frac{\sigma_{\rm
    MS}^2}{\sin^2{\vartheta_C}}+\sigma_{\Phi_{hit}}^2 .
\quad
 \nonumber
\end{equation}
The hit-related uncertainties $\sigma_{\theta_{hit}}$ and
$\sigma_{\Phi_{hit}}$ are obtained by propagating the uncertainties
from the hit positions, see Appendix~\ref{appendix_kink}.
Note that this simplified procedure does not fit the hit positions,  i.e. ${\bf x_k}={\bf
  \xi_k}$,
but it yields the correct $\chi^2$-distribution, i.e. the probability distribution
is flat. This is essential to validate the hit triplets for the track reconstruction. 

\begin{figure}[tb]
\centering
\includegraphics[width=0.75\textwidth]{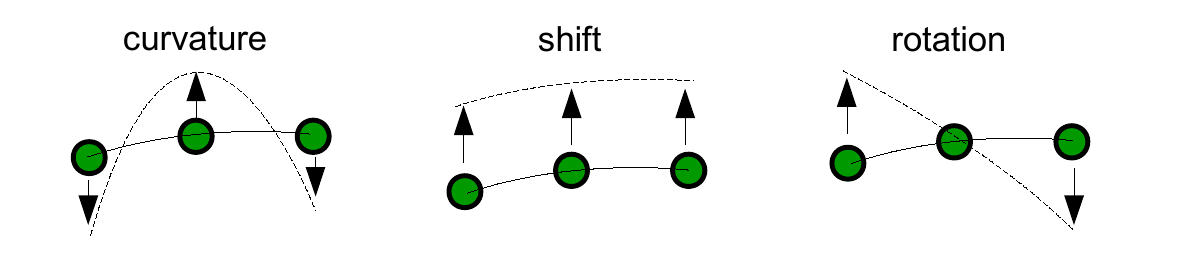}
\caption{Representation of hit uncertainty modes of a hit triplet. 
  From left to right:
  curvature uncertainty, shift uncertainty and rotation uncertainty.
  }
\label{fig:sketch_triplet_degrees}
\end{figure}

Finally,
the three-dimensional radius is obtained by calculating:
\begin{equation}
R_{\rm{3D}}^{min}   =  - \; \frac{\eta \; \tilde{\Phi}_{\rm C} \sin{\vartheta_C}^2
  /\sigma_{\phi}(R_{\rm{3D}}^{min})^2  
+ \beta \; \tilde{\Theta}_{\rm C} /\sigma_{\theta}(R_{\rm{3D}}^{min})^2
    }{\eta^2 \sin{\vartheta_C}^2/\sigma_{\phi}(R_{\rm{3D}}^{min})^2 + \beta^2/\sigma_{\theta}(R_{\rm{3D}}^{min})^2}
 .
 \nonumber
\end{equation}
By using the Highland approximation (Equation~\ref{eq:highland})
above equation
can be transformed into a cubic equation in
$R_{\rm{3D}}^{min}$ which is algebraically solvable, see Appendix~\ref{app:hit_uncertainties}.

The inclusion of the hit uncertainties generates nine extra degrees of freedom.
Relevant for track fitting and reconstruction are only six\footnote{
The other three degrees of freedom correspond to two oscillation modes and a
shift in longitudinal direction, which are completely negligible because their small size.
} ,
which can be represented as ``triplet bending'', ``triplet shifting'' and ``triplet
rotation'' in the transverse as well as in the longitudinal plane, 
see Figure~\ref{fig:sketch_triplet_degrees}.
The triplet bending is included in above fitting
procedure, whereas shifts and rotations of the hit triplet are not. 
For track extrapolations, however, they might be relevant and have to be included.
The rotational uncertainty of a hit triplet is given in Appendix~\ref{app:rotation}.

\section{Triplet Linking and Merging}
\label{sec:merging}
For the linking of hit triplets (or between a
validated triplet and a track prototype)
two strategies can be employed, the intersection technique
and the extrapolation technique. 
In the intersection technique hits shared between validated triplets  are used to identify connected tracks,
 see Figure~\ref{fig:triplet_merging}.
This can be done quickly in hardware by using fast lookup techniques or in software by
using binary search trees, for example.

The extrapolation technique extrapolates already reconstructed track prototypes
or triplets into other detector layers to add new single hits or hit triplets. 
This method is very efficient if the search window is small. That is
typically the case for long track prototypes for which the track parameters have
been precisely fitted in the previous step.
Both techniques can also be combined to speed up linking. 

The consistency of a new track hypothesis
can be tested by calculating a combined
$\chi^2_{comb}$ value by summing over all contributing triplets, 
where the number of hit triplets given is by $n_{hit}-2$.
There is some freedom in how to perform the weighting in the sum. 
For dominating multiple scattering uncertainties and 
if the $R_{\rm{3D},i}$ were calculated using Equation~\ref{eq:r3d}
the optimal sum reads:
\begin{equation}
\chi_{comb}^2  = 
\sum_{i}^{n_{hit}-2}  \chi_i^2 \ +
\  \frac{(R_{\rm{3D},i}-\overline{R_{\rm{3D}}})^2}{\sigma_i(\overline{R_{3D}})^2}
\label{eq:chi2comb}
\end{equation}
with
\begin{equation}
	\overline{R_{3D}} = \sum_{i}^{n_{hit}-2}
        \frac{R_{3D,i}^2}{\sigma(R_{3D,i})^2}
 \; \big/ \; 
 \sum_{i}^{n_{hit}-2}        \frac{R_{3D,i}}{\sigma(R_{3D,i})^2} .
\label{eq:r3d_average}
\end{equation}
This definition of the averaged three-dimensional radius is free of bias.
Equations~\ref{eq:chi2comb} and~\ref{eq:r3d_average} are exact and no refitting 
of the track parameters after the linking step is required.

For dominating hit uncertainties, the relative momentum resolution is
no longer constant and it is advantageous to perform 
the $\chi^2$ sum and 
the averaging in three-dimensional curvature, 
$h_{3D}=1/R_{3D}$, yielding:
\begin{equation}
\chi_{comb}^2  = 
\sum_{i}^{n_{hit}-2}  \chi_i^2 \ +
\  \frac{(h_{3D,i}-\overline{h_{3D}})^2}{\sigma_i(\overline{h_{3D}})^2}
\label{eq:h3d_chi2comb}
\end{equation}
with
\begin{equation}
	\overline{h_{3D}} = \sum_{i}^{n_{hit}-2}
        \frac{h_{3D,i}}{\sigma(h_{3D,i})^2} 
\; \big/ \; 
\sum_{i}^{n_{hit}-2}        \frac{1}{\sigma(h_{3D,i})^2}   .
\label{eq:h3d_average}
\end{equation}
Equations~\ref{eq:h3d_chi2comb} 
and~\ref{eq:h3d_average} are not exact 
since long range correlations are ignored and triplet parameter uncertainties
are overestimated
in the procedure described in Section~\ref{sec:triplet_general}.
The so obtained combined $\chi^2$ value is only a lower (optimistic) estimate.
However, this combination procedure is
very fast, highly parallelizable and useful for track fitting as it allows to quickly reject
wrong triplet combinations without refitting.

The full precision is finally obtained by fitting, in addition to the kink angles, the
hit positions ${\bf x}$, see next section.

%

\section{General Broken Line Fit}
\label{sec:gbl}

\begin{figure}[t!]
\centering 
  \includegraphics[width=0.65\textwidth]{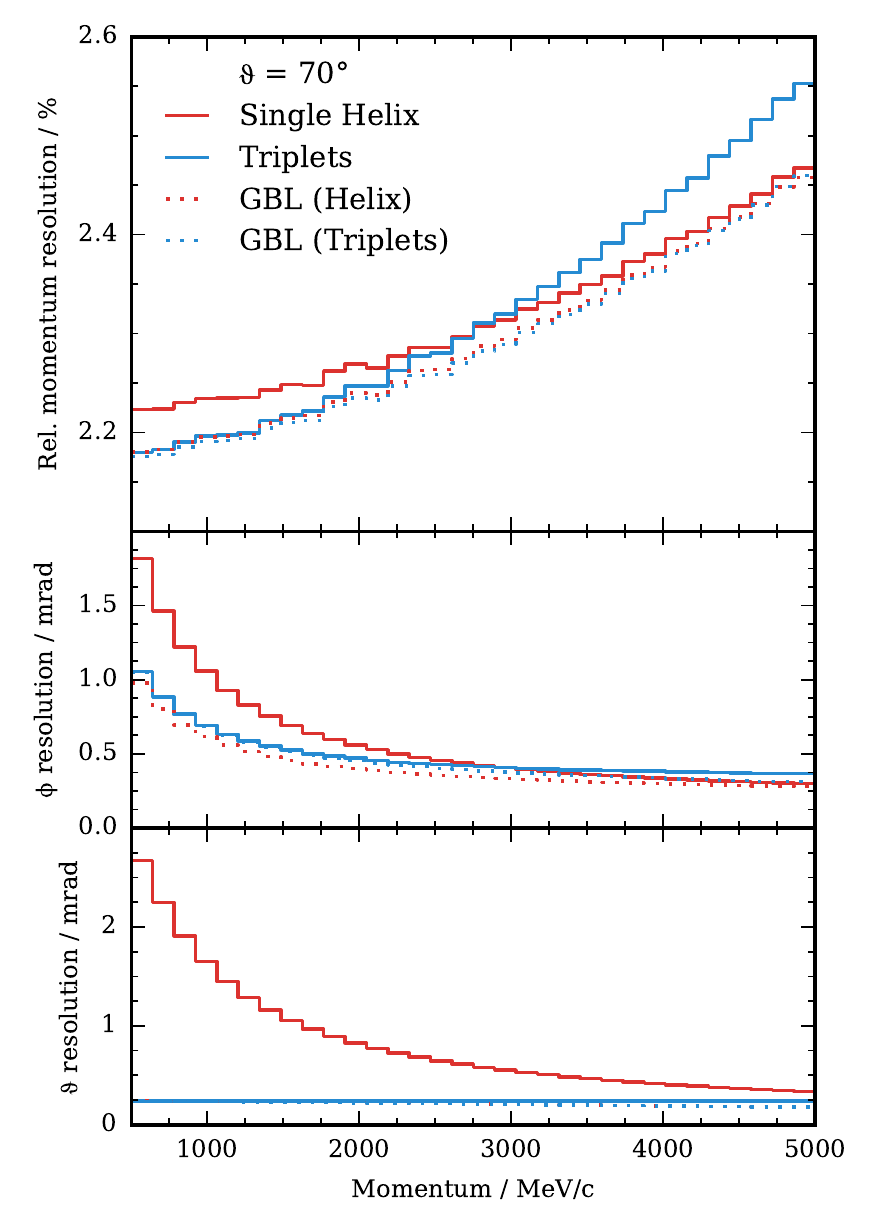}
  \caption{Track parameter resolution for an equidistant layer geometry
    of a high energy LHC-type detector as described in \cite{ref:triplet_paper}.
    The top panel shows
    the relative momentum resolution and the bottom panels show the
    resolution of the azimuthal angle $\phi$ and the polar
    angle $\vartheta$ for tracks with a
    polar angle of $\vartheta=70^\circ$. 
    Plot taken from~\cite{ref:triplet_paper}.
  }
\label{fig:track_resolution}
\end{figure}

Long-range correlations among different hit triplets can be included by the GBL
fit~\cite{Blobel:2006:nft, BlobelEtAl:2011:fac, Kleinwort:2012:gbl}.
It performs a linearization around a reference trajectory which is here chosen
to be the broken line solution discussed in the previous section.
A GBL fit using the triplets fit as reference has been recently implemented
in~\cite{Kiehn2016,Watson}.

The performance of this fit and the improvement with respect to the triplets fit
solution and a solution obtained from a single helix fit
has been studied in~\cite{ref:triplet_paper}
for a typical high energy LHC detector with equidistant layer spacing.
The track parameter resolutions obtained by the different track fits for
particles at $\vartheta=70^\circ$ are shown in 
Figure~\ref{fig:track_resolution} for the momentum, azimuthal angle and polar angle.
In the momentum range up to about $2$~GeV/c the triplets fit provides a
resolution which is equally good as the GBL fits. 
In this region the triplet fit is considered to be adequate. 
Note that more than 99\% of all charged particles produced at the LHC
have a transverse momentum below $1$~GeV. 

In the momentum range of above $2$~GeV/c the track parameter resolutions is
significantly improved by using the GBL fit. 
For even higher momenta $p\ge 5$~GeV/c a single helix fit can be applied
which requires far less computational effort than the GBL fit.
The combination of the triplets fit for low momenta and a single helix fit for
high momenta provide a very efficient scheme for fast track reconstruction.
Only in the intermediate regime between about $2$ to $5$~GeV/c significant
performance improvements are possible by using the GBL fit.

The triplet based track reconstruction method has been successfully
implemented for the 
Mu3e-experiment~\cite{RP,Berger:2013raa} and tested in simulations. 
The Mu3e-experiment will search for the
lepton flavor violating decay of stopped muons $\mu^+ \rightarrow e~^+ e~^+ e~^-$.
Because of the very low momentum of the decay electrons tracking is performed
in  kinematic region which is dominated by multiple scattering and spatial
hit uncertainties can be ignored. 
For the Mu3e-experiment the above described track reconstruction based on hit
triplets was successfully implemented on GPUs.

Triplet based tracking was also implemented 
for an upgraded ATLAS detector designed for High Luminosity LHC. 
It was successfully tested
for different detector geometries and different layer spacings.
A publication of the results is foreseen in the near future.

\section{Summary and Outlook}
A new track reconstruction method based on hit triplets has been proposed.
It is based on a multiple scattering based track
model~\cite{ref:triplet_paper}
which has been extended here to include also spatial hit uncertainties.
The described triplet fit has an algebraic solution and can be used
to quickly validate hit combinations. 
It is therefore ideally suited for track reconstruction in general, 
and track seed searches in particular.
The proposed method gives a simple prescription for the combination of hit triplets.
The fit quality and the track momentum can be represented by sums, which can
be quickly calculated. 
This method can be easily combined with a General Broken Line Fit 
to improve the track parameters for non-vanishing spatial hit uncertainties.
For high momentum tracks (negligible multiple scattering) 
a single helix fit can be used alternatively, which
is much faster than the GBL fit.

The whole track reconstruction procedure is non-iterative. All hit triplets
can be reconstructed in parallel, what makes it ideal for the implementation 
on parallel computing architectures.

The used tracking model relies on the assumption that the magnetic field
within a triplet is homogeneous and that multiple scattering occurs only at or
near the detection layers. The effect of non-homogeneous magnetic fields, 
particles with missing hits in layers due to detector efficiencies,
and particle scattering in material
outside the detection layers will be subject of further studies.

\section*{Acknowledgments}
The author thanks Moritz~Kiehn for helping to implement the GBL fit and Frank~Meier-Aeschbacher
for fruitful discussions and reading of the manuscript.

\newpage
\appendix

\section{Propagation of Triplet Hit Uncertainties}
\label{appendix_kink}
\begin{figure}[t!]
\centering 
  \includegraphics[width=0.7\textwidth]{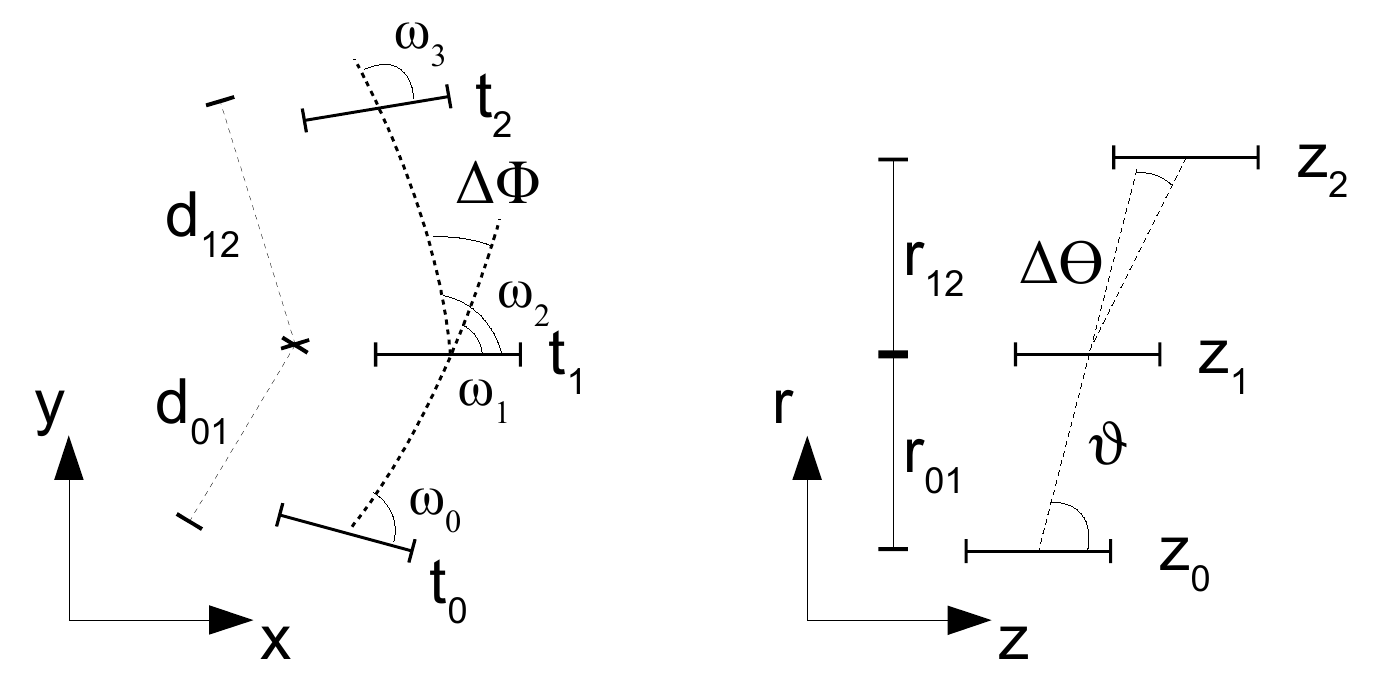}
  \caption{Hit triplet with spatial hit uncertainties and definition of the 
    kink angle uncertainties in the transverse (left)
    and longitudinal (right) plane. 
  }
\label{fig:sketch_hit_unc}
\end{figure}
A simple method to include hit uncertainties in the triplet fit is to interpret
the hit residuals as kink angles: $\Delta \Phi$ in the transverse
and $\Delta \Theta$ in the longitudinal
plane, see Figure~\ref{fig:sketch_hit_unc}.
The tilt angles  with respect to the particle trajectory are given by the
angles $\omega_i$. 
The uncertainties in radial direction ${\sigma_{r_k}}$, in azimuthal
direction ${\sigma_{t_k}}$ and in axial direction ${\sigma_{z_k}}$ can be
propagated to define effective kink angle uncertainties  $\sigma_{\phi_{hit}}$ and 
$\sigma_{\theta_{hit}}$:
\begin{align}
\sigma_{\phi_{hit}}^2 & =  
\frac{\sin^2{\omega_0}}{\Delta d_{01}^2} \; \sigma_{t_0}^2 \; + \;
\frac{\sin^2{\omega_3}}{\Delta d_{12}^2} \; \sigma_{t_2}^2 \; + \;
\left(\frac{\sin{\omega_1}}{\Delta d_{01}} + \frac{\sin{\omega_2}}{\Delta d_{12}} \right)^2  \sigma_{t_1}^2
\nonumber \\
& +   
\frac{\cos^2{\omega_0}}{\Delta d_{01}^2} \; \sigma_{r_0}^2 \; + \;
\frac{\cos^2{\omega_3}}{\Delta d_{12}^2} \; \sigma_{r_2}^2 \; + \;
\left(\frac{\cos{\omega_1}}{\Delta d_{01}} + \frac{\cos{\omega_2}}{\Delta d_{12}} \right)^2  \sigma_{r_1}^2
\end{align}
\begin{align}
\sigma_{\theta_{hit}}^2 & =  
\sin^4{\vartheta} \; \left( 
\frac{\sigma_{z_0}^2}{\Delta r_{01}^2} + \frac{\sigma_{z_2}^2}{\Delta r_{12}^2} + \sigma_{z_1}^2 \;
\left(\frac{1}{\Delta r_{01}} + \frac{1}{\Delta r_{12}} \right)^2 \right)
\nonumber \\ 
& +  
\sin^2{\vartheta} \; \cos^2{\vartheta} \; \left( 
\frac{\sigma_{r_0}^2}{\Delta r_{01}^2} + \frac{\sigma_{r_2}^2}{\Delta r_{12}^2} + \sigma_{r_1}^2 \;
\left(\frac{1}{\Delta r_{01}} + \frac{1}{\Delta r_{12}} \right)^2 \right)  \ .
 \nonumber
\label{eq:sigma_theta_hit}
\end{align}

\section{Triplet Rotation Uncertainty}
\label{app:rotation}
For non-vanishing spatial hit uncertainties rotational degrees of freedom
in the transverse and in the longitudinal plane are generated.
These rotational uncertainties are important for track extrapolations.
For a barrel-type detector and negligible track curvatures the rotational 
uncertainties are given by:
\begin{align}
\delta \theta_{rot}^2 & = \sin^4{\vartheta} \;  \; 
\frac{\sigma_{z_0}^2 \sigma_{z_1}^2 + \sigma_{z_1}^2 \sigma_{z_2}^2 + \sigma_{z_0}^2 \sigma_{z_2}^2}
{\sigma_{z_0}^2 r_{12}^2 + \sigma_{z_2}^2 r_{01}^2 +
\sigma_{z_1}^2 r_{02}^2} 
 \nonumber \\
\delta \phi_{rot}^2 & =  
\frac{\sigma_{t_0}^2 \sigma_{t_1}^2 + \sigma_{t_1}^2 \sigma_{t_2}^2 + \sigma_{t_0}^2 \sigma_{t_2}^2}
{\sigma_{t_0}^2 d_{12}^2 + \sigma_{t_2}^2 d_{01}^2 +
\sigma_{t_1}^2 d_{02}^2} .
 \nonumber
\end{align}
Here the same notation as in Appendix~\ref{appendix_kink} is used.
It is straight-forward to calculate the rotational uncertainties for 
other detector orientations.

\section{Extended Triplet Fit with Hit Uncertainties}
\label{app:hit_uncertainties}
The three-dimensional radius in an extended triplet fit including hit
uncertainties is given by:
\begin{align}
R_{{3D}}^{min}  & =  - \; \frac{\eta \; \tilde{\Phi}_{ C} \sin{\vartheta}^2
  /\sigma_{\phi}(R_{{3D}}^{min})^2  
+ \beta \; \tilde{\Theta}_{C} /\sigma_{\theta}(R_{{3D}}^{min})^2
    }{\eta^2 \sin{\vartheta}^2/\sigma_{\phi}(R_{{3D}}^{min})^2 + \beta^2/\sigma_{\theta}(R_{{3D}}^{min})^2}
 .
 \nonumber
\end{align}
The polar angle of the track $\vartheta$ can be taken from the circle
solution, which also can serve as a first estimate to calculate the
inclination angles $\omega_i$ in  the transverse plane, see Figure~\ref{fig:sketch_hit_unc}.
Note that in general $\sigma_{\phi}\ne \sigma_{\theta}$.

A difficulty arises from the fact 
that the multiple scattering uncertainty depends on the particle momentum,
which is not known before track fitting.
A solution is to include this momentum dependence in the fit
by using the Highland parameterization (Equation~\ref{eq:highland}), 
leading to the following cubic equation: 
\begin{align}
0 
& = 
{R_{{3D}}^{min}}^3 \; 
(\eta^2 \sigma_{\theta_{hit}}^2  + \beta^2   \sigma_{\phi_{hit}}^2)  \nonumber \\
& + 
{R_{{3D}}^{min}}^2 \; 
(\eta \; \tilde{\Phi}_{ C} \; \sigma_{\theta_{hit}}^2 + 
\beta \; \tilde{\Theta}_{ C} \; \sigma_{\phi_{hit}}^2)
 \nonumber \\
& +
R_{{3D}}^{min} \; b^2\;
  (\eta^2+\beta^2 / \sin{\vartheta}^2)  \nonumber \\
&+  
b^2\;
(\eta \; \tilde{\Phi}_{ C} + \beta \; \tilde{\Theta}_{C} /
\sin{\vartheta}^2)
 \nonumber \\
& :=  a_3 {R_{{3D}}^{min}}^3 + a_2 {R_{{3D}}^{min}}^2 + a_1
{R_{{3D}}^{min}} + a_0 
  .
 \nonumber
\end{align}
Here $b$ is an effective scattering parameter, which is a measure for the
material thickness at the scattering layer, see~\cite{ref:triplet_paper} for definition. 
Using Cardano's formula it is straight forward to find the solution for
$R_{{3D}}^{min}$.

\bibliography{references}
%
%
%
%

\end{document}